# The Calculation of Clebsh-Gordan Coefficients for the Permutation Group by the Eigenfunction Method

Chin-Sheng Wu
Center for General Education
Yuan-Ze University, Taiwan

**Abstract**. We use the eigenfunction method to calculate the Clebsh-Gordan coefficients for the permutation group $S_3$. This method is well-established by Jin-Quan Chen. Here we elaborate the detailed procedures for the pedagogical purpose. Due to the nature of the symmetry, one may get the degeneracy from the solution of eigenfunctions for given one class operator. In order to remove the degeneracy we use extra class operators, which may be the subgroup class operator or even the state permutation operator. In doing so, a variety of eigenvalues come out. Every eigenfunction is therefore obtained, and basis vectors are completely found.

## I. INTRODUCTION

The group theory has been applied extensively in atomic spectroscopy, nuclear physics and elementary particles. In the application, the basis vectors are traditionally represented in terms of wave functions[1,2]. These basis vectors constitute new irreducible representations for their coupling spectra through inner or outer product[3-5]. We therefore elaborate some examples to understand this field.

## II. CALCULATION

We start out with the trivial permutation group $S_2$. Suppose that there are two particles 1 and 2, and two coordinates $\alpha$ and $\beta$. There are two quantum states

$$\varphi_1 = |\alpha\beta\rangle = \chi_\alpha(1)\chi_\beta(2), \tag{1}$$

$$\varphi_2 = |\beta\alpha\rangle = \chi_\beta(1)\chi_\alpha(2), \tag{2}$$

which form a reducible representation of $S_2$. The class operator of $S_2$ is $C$ = (1 2).

$$C_{ij} = \langle \varphi_i | C | \varphi_j \rangle. \tag{3}$$

The group representation of $C$ is therefore

$$\begin{bmatrix} 0 & 1 \\ 1 & 0 \end{bmatrix}. \tag{4}$$

The eigenvalues of $C$ are 1 and -1, which are corresponding to two irreducible representations. In this case of $S_2$, each irrep (irreducible representation) has one dimension. One dimension is related to one basis vector. Two basis vectors are

$$\psi^{(1)} = \sqrt{\frac{1}{2}}(\varphi_1 + \varphi_2), \quad \psi^{(-1)} = \sqrt{\frac{1}{2}}(\varphi_1 - \varphi_2). \tag{5}$$

The Clebsch-Gordan coefficients are directly read from the above equations. Traditionally we use Young tableau to label the irreducible representations

$$\left|\begin{matrix} 1 & 2 \end{matrix}\right\rangle = \psi^{(1)}, \quad \left|\begin{matrix} 1 \\ 2 \end{matrix}\right\rangle = \psi^{(-1)}. \tag{6}$$

These two irreps constitute a reducible representation of $S_2$.

We next consider the permutation group $S_3$ in the configuration $\alpha^2 \beta$. There are three particles in three possible states $\varphi_1 = |\alpha\alpha\beta\rangle$, $\varphi_2 = |\alpha\beta\alpha\rangle$, and $\varphi_3 = |\beta\alpha\alpha\rangle$. Group representations[1,2] of (12), (23) and (13) can be found as

$$(12) \begin{bmatrix} \varphi_1 \\ \varphi_2 \\ \varphi_3 \end{bmatrix} = \begin{bmatrix} 1 & 0 & 0 \\ 0 & 0 & 1 \\ 0 & 1 & 0 \end{bmatrix} \begin{bmatrix} \varphi_1 \\ \varphi_2 \\ \varphi_3 \end{bmatrix}, \tag{7}$$

$$(23) \begin{bmatrix} \varphi_1 \\ \varphi_2 \\ \varphi_3 \end{bmatrix} = \begin{bmatrix} 0 & 1 & 0 \\ 1 & 0 & 0 \\ 0 & 0 & 0 \end{bmatrix} \begin{bmatrix} \varphi_1 \\ \varphi_2 \\ \varphi_3 \end{bmatrix}, \tag{8}$$

$$(13) \begin{bmatrix} \varphi_1 \\ \varphi_2 \\ \varphi_3 \end{bmatrix} = \begin{bmatrix} 0 & 0 & 1 \\ 0 & 1 & 0 \\ 1 & 0 & 0 \end{bmatrix} \begin{bmatrix} \varphi_1 \\ \varphi_2 \\ \varphi_3 \end{bmatrix}, \tag{9}$$

The class operator is $C(3) = (12) + (23) + (13)$. The corresponding eigenequation is

$$\begin{bmatrix} 1 & 1 & 1 \\ 1 & 1 & 1 \\ 1 & 1 & 1 \end{bmatrix} \begin{bmatrix} u_1 \\ u_2 \\ u_3 \end{bmatrix} = v \begin{bmatrix} u_1 \\ u_2 \\ u_3 \end{bmatrix}, \tag{10}$$

where $\varphi^{(v)} = u_1\varphi_1 + u_2\varphi_2 + u_3\varphi_3$.

The eigenvalues are $v = 3, 0$, and $0$.
When $v = 3$, $u_1 = u_2 = u_3$.

When $v = 0$, $u_1 + u_2 + u_3 = 0$.

We need the subgroup $S_2 \subset S_3$ basis to solve the above equation completely. $C(2) = (12)$ is the class operator of $S_2$. The solutions to the eigenequation are
$m = 1$, $u_2 = u_3$, which is the double root,
$m = -1$, $u_1 = 0$, $u_2 = -u_3$.

We take the simultaneous solution to the following eigenequation

$$\begin{pmatrix} C(3) \\ C(2) \end{pmatrix} \psi_{(m)}^{(v)} = \begin{pmatrix} v \\ m \end{pmatrix} \psi_{(m)}^{(v)}. \tag{11}$$

$(v, m) = (3, 1) \quad u_1 = u_2 = u_3$,
$(v, m) = (0, 1) \quad u_1 = -2u_2 = -2u_3$,
$(v, m) = (0, -1) \quad u_1 = 0 \quad u_2 = -u_3$.

The basis vectors are

$$\psi_1^{(3)} = |1 \ 2 \ 3\rangle = \sqrt{\frac{1}{3}}(\varphi_1 + \varphi_2 + \varphi_3), \tag{12}$$

$$\psi_1^{(0)} = \left|\begin{matrix} 1 & 2 \\ 3 & \end{matrix}\right\rangle = \sqrt{\frac{1}{6}}(2\varphi_1 - \varphi_2 - \varphi_3), \tag{13}$$

$$\psi_{-1}^{(0)} = \left|\begin{matrix} 1 & 3 \\ 2 & \end{matrix}\right\rangle = \sqrt{\frac{1}{2}}(\varphi_2 - \varphi_3). \tag{14}$$

Finally we consider the permutation group $S_3$ in the configuration $\alpha\beta\gamma$. There are three particles in the six possible states $|\alpha\beta\gamma\rangle, |\beta\alpha\gamma\rangle, |\gamma\beta\alpha\rangle, |\alpha\gamma\beta\rangle, |\gamma\alpha\beta\rangle, |\beta\gamma\alpha\rangle$

$$\begin{pmatrix} C(3) \\ C(2) \\ C'(2) \end{pmatrix} \psi_m^{(v)\ k} = \begin{pmatrix} v \\ m \\ k \end{pmatrix} \psi_m^{(v)\ k}, \tag{15}$$

$$\psi_m^{(v)\ k} = \sum_{i=1}^{6} u_i \varphi_i. \tag{16}$$

The group representations of $C(12)$, $C(23)$ and $C(13)$ are

$$(12)\begin{bmatrix} u_1 \\ u_2 \\ u_3 \\ u_4 \\ u_5 \\ u_6 \end{bmatrix} = \begin{bmatrix} 0 & 1 & 0 & 0 & 0 & 0 \\ 1 & 0 & 0 & 0 & 0 & 0 \\ 0 & 0 & 0 & 0 & 1 & 0 \\ 0 & 0 & 0 & 0 & 0 & 1 \\ 0 & 0 & 1 & 0 & 0 & 0 \\ 0 & 0 & 0 & 1 & 0 & 0 \end{bmatrix} \begin{bmatrix} u_1 \\ u_2 \\ u_3 \\ u_4 \\ u_5 \\ u_6 \end{bmatrix}, \tag{17}$$

$$(23)\begin{bmatrix} u_1 \\ u_2 \\ u_3 \\ u_4 \\ u_5 \\ u_6 \end{bmatrix} = \begin{bmatrix} 0 & 0 & 0 & 1 & 0 & 0 \\ 0 & 0 & 0 & 0 & 0 & 1 \\ 0 & 0 & 0 & 0 & 1 & 0 \\ 1 & 0 & 0 & 0 & 0 & 0 \\ 0 & 0 & 1 & 0 & 0 & 0 \\ 0 & 1 & 0 & 0 & 0 & 0 \end{bmatrix} \begin{bmatrix} u_1 \\ u_2 \\ u_3 \\ u_4 \\ u_5 \\ u_6 \end{bmatrix}, \tag{18}$$

$$(13)\begin{bmatrix} u_1 \\ u_2 \\ u_3 \\ u_4 \\ u_5 \\ u_6 \end{bmatrix} = \begin{bmatrix} 0 & 0 & 1 & 0 & 0 & 0 \\ 0 & 0 & 0 & 0 & 1 & 0 \\ 1 & 0 & 0 & 0 & 0 & 0 \\ 0 & 0 & 0 & 0 & 0 & 1 \\ 0 & 1 & 0 & 0 & 0 & 0 \\ 0 & 0 & 0 & 1 & 0 & 0 \end{bmatrix} \begin{bmatrix} u_1 \\ u_2 \\ u_3 \\ u_4 \\ u_5 \\ u_6 \end{bmatrix}. \tag{19}$$

Let $C(3) = (12) + (23) + (13)$. The group representation is therefore

$$\begin{bmatrix} -v & 1 & 1 & 1 & 0 & 0 \\ 1 & -v & 0 & 0 & 1 & 1 \\ 1 & 0 & -v & 0 & 1 & 1 \\ 1 & 0 & 0 & -v & 1 & 1 \\ 0 & 1 & 1 & 1 & -v & 0 \\ 0 & 1 & 1 & 1 & 0 & -v \end{bmatrix} \begin{bmatrix} u_1 \\ u_2 \\ u_3 \\ u_4 \\ u_5 \\ u_6 \end{bmatrix} = 0. \tag{20}$$

Using Laplace development for Cramer's rule, one gets

$$= -v \begin{vmatrix} -v & 0 & 0 & 1 & 1 \\ 0 & -v & 0 & 1 & 1 \\ 0 & 0 & -v & 1 & 1 \\ 1 & 1 & 1 & -v & 0 \\ 1 & 1 & 1 & 0 & -v \end{vmatrix} - \begin{vmatrix} 1 & 0 & 0 & 1 & 1 \\ 1 & -v & 0 & 1 & 1 \\ 1 & 0 & -v & 1 & 1 \\ 0 & 1 & 1 & -v & 0 \\ 0 & 1 & 1 & 0 & -v \end{vmatrix}$$

$$+ \begin{vmatrix} 1 & -v & 0 & 1 & 1 \\ 1 & 0 & 0 & 1 & 1 \\ 1 & 0 & -v & 1 & 1 \\ 0 & 1 & 1 & -v & 0 \\ 0 & 1 & 1 & 0 & -v \end{vmatrix} - \begin{vmatrix} 1 & -v & 0 & 1 & 1 \\ 1 & 0 & -v & 1 & 1 \\ 1 & 0 & 0 & 1 & 1 \\ 0 & 1 & 1 & -v & 0 \\ 0 & 1 & 1 & 0 & v \end{vmatrix}$$

$$= -v(v^3(6-v^2)) - v^4 - v^4 - v^4$$
$$= v^4(v^2 - 9)$$
$$= 0 \tag{21}$$

$v = 3: u_2 = u_3 = u_4, u_1 = u_5 = u_6$
$v = -3: u_1 = u_2 = u_3 = u_4 = u_5 = u_6$
$v = 0: u_1 + u_5 + u_6 = 0, u_2 + u_3 + u_4 = 0$

Similar to the previous example we have subgroup operator $C(2) = (12)$.
The group representation is

$$\begin{bmatrix} -m & 1 & 0 & 0 & 0 & 0 \\ 1 & -m & 0 & 0 & 0 & 0 \\ 0 & 0 & -m & 0 & 0 & 1 \\ 0 & 0 & 0 & -m & 1 & 0 \\ 0 & 0 & 0 & 1 & -m & 0 \\ 0 & 0 & 1 & 0 & 0 & -m \end{bmatrix} \begin{bmatrix} u_1 \\ u_2 \\ u_3 \\ u_4 \\ u_5 \\ u_6 \end{bmatrix} = 0. \tag{22}$$

Using extra state permutation operator $C'(2)^1$, we take the simultaneous solution for the following eigenequation

$$\begin{pmatrix} C(3) \\ C(2) \\ C'(2) \end{pmatrix} \psi_m^{(v)\ k} = \begin{pmatrix} v \\ m \\ k \end{pmatrix} \psi_m^{(v)\ k}, \tag{23}$$

and get $m = \pm 1: u_1 = \pm u_2, u_3 = \pm u_6, u_4 = \pm u_5$

$k = \pm 1: u_1 = \pm u_2, u_3 = \pm u_5, u_4 = \pm u_6$

$(v, m, k) = (0,1,1): u_1 = u_2 = 2u_3 = 2u_4 = 2u_5 = 2u_6$
$(0,1,-1): u_1 = u_2 = 0, u_3 = -u_4 = -u_5 = u_6$
$(0,-1,1): u_1 = u_2 = 0, -u_3 = u_4 = -u_5 = u_6$
$(0,-1,-1): u_1 = -u_2 = 2u_3 = 2u_4 = -2u_5 = -2u_6$

The basis vectors are

$$\psi_1^{(0)\ 1} = \begin{vmatrix} 1 & 2 & \alpha & \beta \\ 3 & & & \gamma \end{vmatrix} = \sqrt{\frac{1}{12}}[2(\varphi_1 + \varphi_2) - (\varphi_3 + \varphi_4 + \varphi_5 + \varphi_6)] \tag{24}$$

$$\psi_{-1}^{(0)\ 1} = \begin{vmatrix} 1 & 3 & \alpha & \beta \\ 2 & & & \gamma \end{vmatrix} = \frac{1}{2}(-\varphi_3 + \varphi_4 - \varphi_5 + \varphi_6) \tag{25}$$

$$\psi_1^{(0)\ -1} = \begin{vmatrix} 1 & 2 & \alpha & \gamma \\ 3 & & & \beta \end{vmatrix} = \frac{1}{2}(-\varphi_3 + \varphi_4 + \varphi_5 - \varphi_6) \tag{26}$$

$$\psi_{-1}^{(0)\ -1} = \begin{vmatrix} 1 & 3 & \alpha & \gamma \\ 2 & & & \beta \end{vmatrix} = \sqrt{\frac{1}{12}}[2(\varphi_1 - \varphi_2) + \varphi_3 + \varphi_4 - \varphi_5 - \varphi_6)] \tag{27}$$

$$\psi_1^{(3)\ 1} = |1 \quad 2 \quad 3 \quad \alpha \quad \beta \quad \gamma\rangle = \sqrt{\frac{1}{6}}(\varphi_1 + \varphi_2 + \varphi_3 + \varphi_4 + \varphi_5 + \varphi_6) \tag{28}$$

$$\psi_{-1}^{(-3)\ -1} = \begin{vmatrix} 1 & \alpha \\ 2 & \beta \\ 3 & \gamma \end{vmatrix} = \sqrt{\frac{1}{6}}(\varphi_1 - \varphi_2 - \varphi_3 - \varphi_4 + \varphi_5 + \varphi_6). \tag{29}$$

### III. CONCLUDING REMARK

Due to the nature of the symmetry, one may get the degeneracy from the solution of eigenfunctions for given one class operator. In order to remove the degeneracy we use extra class operators, which may be the subgroup class operator or even the state permutation operator. In doing so, a variety of eigenvalues come out. Every eigenfunction is therefore obtained, and basis vectors are completely found.